\documentclass[review]{elsarticle}
\usepackage{multirow}

\usepackage{graphicx, amsmath, breqn}
\usepackage{bm}
\usepackage{color}

\begin{document}

\begin{frontmatter}

\title{Degree correlations in signed social networks}

\author{Valerio Ciotti}
\address{School of Business and Management \& School of Mathematical Sciences, Queen Mary University of London, Mile End Road, E1 4NS London, UK}
\author{Ginestra Bianconi}
\address{School of Mathematical Sciences, Queen Mary University of London, Mile End Road, E1 4NS London, UK}
\author{Andrea Capocci}
\address{Institute for Complex Systems - CNR, Unit of Sapienza University \& Department of Physics, Sapienza University, Piazzale Aldo Moro 5, 00185 Rome, Italy}
\author{Francesca Colaiori}
\address{Institute for Complex Systems - CNR, Unit of Sapienza University \& Department of Physics, Sapienza University, Piazzale Aldo Moro 5, 00185 Rome, Italy}
\author{Pietro Panzarasa}
\address{School of Business and Management, Queen Mary University of London, Mile End Road, E1 4NS London, UK}





\begin{abstract}
We investigate degree correlations in two online social networks where users are connected through different types of links. We find that, while subnetworks in which links have a positive connotation, such as endorsement and trust, are characterized by assortative mixing by degree, networks in which links have a negative connotation, such as disapproval and distrust, are characterized by disassortative patterns. We introduce a class of simple theoretical models to analyze the interplay between network topology and the superimposed structure based on the sign of links. Results uncover the conditions that underpin the emergence of the patterns observed in the data, namely the assortativity of positive subnetworks and the disassortativity of negative ones. We discuss the implications of our study for the analysis of signed complex networks.   
\end{abstract}

\begin{keyword}
\texttt mixing patterns by degree \sep assortativity \sep disassortativity \sep structural balance \sep signed networks \sep positive and negative subnetworks
\end{keyword}

\end{frontmatter}

\section{Introduction}

Over the last few years, an increasing interest in the study of social networks has prompted physicists, mathematicians and computer scientists to join sociologists in their endeavors to develop network models concerned with the antecedents, structure, and evolution of social interaction \cite{Erdos1959a,Watts1998,Barabasi1999}. Recent studies have indicated that social networks across many empirical domains display the typical signature of complex networks, namely the long-tailed distribution of the degrees of nodes \cite{Barabasi1999}. In addition to this, an attempt has been made to uncover the distinctive structural features and empirical regularities that distinguish social networks from other types of complex networks. While in most real networks degrees of neighboring nodes tend to be anticorrelated, research has suggested that social networks tend to be characterized by the opposite correlation pattern \cite{Newman2002,Newman2003}. The tendency of nodes with similar degree to connect with each other is often referred to as ``assortative mixing by degree'', and has been observed in a number of social networks, including very large-scale online social networks such as Facebook and Twitter \cite{Ugander2011}.

A variety of models have been proposed by physicists, sociologists and computer scientists to explain these distinctive properties of social networks. For instance, assortative mixing has been related to the underlying community structure of social networks \cite{Newman2003}. More recently, assortative mixing has been explained in terms of transitivity \cite{Foster2011}, homophily \cite{Traud2012}, and unsubstitutability of individuals and resources \cite{Holme}. Research has also uncovered distinctive interaction patterns within social signed networks in which relationships can have a positive (e.g., trust and friendship) or negative (e.g., distrust and enmity) connotation \cite{Szell2010}. In particular, the theory of ``structural balance'' has long suggested that, in undirected signed social networks, individuals embedded within closed triads tend to minimize cognitive tension: an individual tends to befriend a friend's friend, distrust a friend's enemy, befriend an enemy's enemy, and distrust an enemy's friend \cite{Cartwright1956, HEIDER1946a}.

Here we focus our attention on the emergence of degree correlations in signed networks, and how these correlations can be used to predict the sign of links in cases where it is not known or cannot be assessed directly. Indeed, despite the ubiquity and salience of negative relationships in a wide range of social systems, the detection of mixing patterns by degree has been confined primarily within the domain of unsigned networks or simply networks in which nodes were assumed to be connected through positive links (e.g., scientific collaboration networks and interlocking directorate networks \cite{Newman2003, Holme}). However, negative networks may exhibit correlation patterns that differ from those detected in positive networks \cite{Szell2010a}. Do individuals who distrust many others tend to distrust each other, or do they channel their negative feelings toward other individuals who distrust only very few others? To address this problem, here we propose a class of simple models that help uncover the relation between the sign of links and the type of degree correlations characterizing a network.  

The outline of the paper is as follows. In Section \ref{data}, we introduce two signed online social networks, and examine the degree distributions and correlations of the positive and negative subnetworks extracted from the data. In Section \ref{model}, we propose a  generative model of signed networks that polarize into two mutually exclusive groups of nodes. Section \ref{random} focuses on the case of random networks with binomial degree distributions, whereas Section \ref{power} deals with more realistic cases of networks with power-law degree distributions. Finally, in Section \ref{three} we extend our modeling framework to networks in which nodes can split into three (or more) hostile groups. In Section \ref{last}, we summarize our findings and discuss their implications for research on signed complex networks. 

\section{The data}
\label{data}
We analyze two online social networks. The first is the network formed by the users of Epinions ({\em www.epinions.com}), a website for user-generated reviews of various products. Registered users of Epinions can declare their trust or distrust toward one another, based on the comments they post. The second social network is formed by the users of Slashdot ({\em www.slashdot.org}), a website devoted to the discussion of technology-related news, and in which the Slashdot Zoo feature enables users to tag one another as ``friends'' or ``foes''. In both Epinions and Slashdot, connections are directed and signed. The meaning of the sign of links is similar: a positive link means that a user endorses another user's comments, whereas a negative one means that a user dislikes another user's comments. Both network datasets are available from the Stanford Network Analysis Project website \cite{SNAP}.

Table \ref{Table1} reports the number of nodes and links in the datasets \cite{Leskovec2010a,Leskovec2010b}. Epinions is composed of $131,828$ nodes and $841,372$ directed links. In particular, $717,667$ of these links (i.e., $85.00\%$) are positive and represent the trust users accord to each other. Moreover, links connecting $130,162$ pairs of nodes in Epinions are reciprocated, of which only $1.8\%$ are characterized by a combination of a positive and a negative sign (i.e., node $i$ points positively to node $j$, and $j$ points negatively to $i$). The Slashdot social network is composed of $82,144$ nodes and $549,202$ links, $425,072$ of which are positive (i.e., $77.40\%$ of the total number of links). Moreover, $48,721$ pairs of nodes are connected through reciprocated links, of which only $4.0\%$ are characterized by different signs. 

\begin{table}
\centering
\begin{tabular}{l||c|c}
 &\multicolumn{1}{|c|}{\textbf{Epinions}} & \multicolumn{1}{|c}{\textbf{Slashdot}}\\ 
\hline
\hline 
\textbf{Nodes} 	& 	131,828		&  82,144    \\ 
\textbf{Links} 	& 841,372		& 549,202    \\ 
\textbf{Positive} &   717,667    & 425,072 \\
& (85.30\%) & (77.40\%)\\
\textbf{Negative}& 123,705 	&  124,130  \\ 
& (14.70\%) & (22.60\%) \\
\textbf{Reciprocated}&    130,162    & 48,721  \\ 
& (15.47\%) & (8.87\%) \\              		
\end{tabular}
\caption{Nodes and links in Epinions and Slashdot.}
\label{Table1}
\end{table}

\begin{figure}
\centering
\includegraphics[scale=0.35]{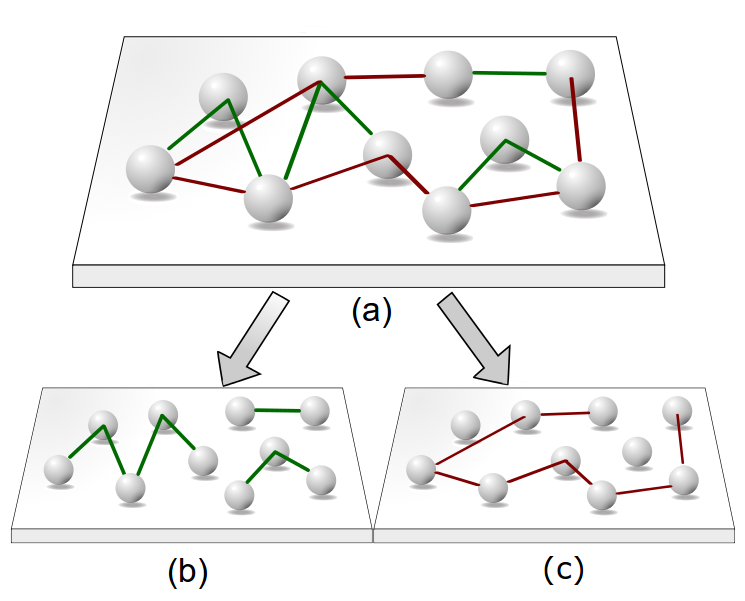}
\caption{Extraction of the positive (b) and negative (c) subnetworks from a signed network (a).}
\label{division}
\end{figure}

To study the impact of the sign of social relationships on the network topology, for each network dataset we filtered out and isolated the positive and the negative subnetworks composed only by reciprocated links of the same sign (see Fig.\ref{division}). In particular, from the Epinions social network two signed subnetworks were extracted: the ``trust'' and ``distrust'' subnetworks in which all links are positive and negative, respectively. Similarly, we created the Slashdot ``friend'' and ``foe'' subnetworks. 

\begin{figure}
\centering
\includegraphics[scale=0.4]{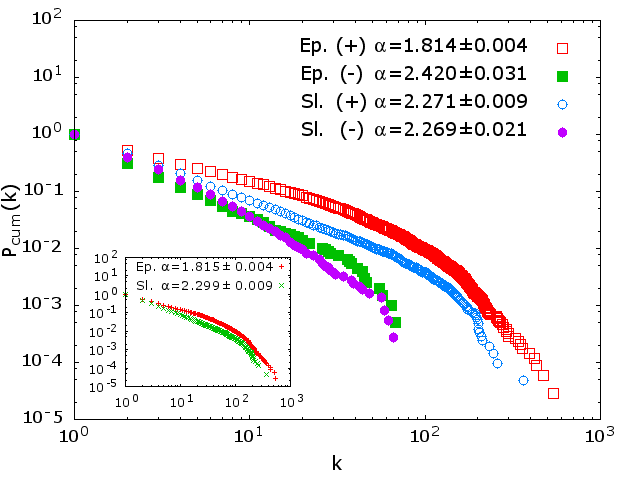} 
\caption{\textbf{Cumulative degree distributions $P_{cum}(k)$ of the Epinions positive (``trust'') and negative (``distrust'') subnetworks and of the Slashdot positive (``friend'') and negative (``foe'') subnetworks}. The inset shows the cumulative degree distributions $P_{cum}(k)$ of the Epinions and Slashdot unsigned networks with reciprocated links. All exponents $\alpha$ of the degree distributions $p(k) \simeq k^{-\alpha}$ have been estimated with the method of maximum likelihood \cite{Alstott}. Kolmogorov-Smirnov tests indicate that all estimated exponents are statistically significant (\textit{p-value} $<0.001$).}
\label{degree}
\end{figure}

These four signed subnetworks are characterized by power-law degree distributions $p(k) \simeq k^{-\alpha}$. Fig.~\ref{degree} shows the cumulative degree distributions $P_{cum}(k)$ of the four subnetworks, with the estimated values of the exponents $\alpha$ of the distributions $p(k)$. The inset of Fig.~\ref{degree} reports the cumulative degree distributions of the unsigned networks with reciprocated links, and the corresponding estimated exponents $\alpha$. 

\subsection{Degree correlations}
Research has typically relied on two fundamental measures for detecting mixing patterns by degree in complex networks. The first measure is the quantity $K_{nn}(k)$, namely the average degree of the nearest neighbors of nodes with degree $k$, defined in \cite{Pastor-Satorras2001} as
\begin{equation}
K_{nn}(k)=\sum_{k'} k' p(k'|k).
\end{equation}
The transitional probability $p(k'|k)$ can be defined as the the conditional probability that a link emanating from a node of degree $k$ is connected to a node of degree $k'$
\begin{equation}
p(k'|k) = \frac{E_{kk'}}{\sum_{k'}{E_{kk'}}} \equiv \frac{ p(k, k')}{ q(k)},
\end{equation}
where $E_{kk'}$ is the entry of the symmetric matrix $E$ that measures the number of links between nodes of degree $k$ and nodes of degree $k'$ for $k \neq k'$, and two times that number for $k=k'$, $p(k, k')$ is the joint probability that a randomly chosen link connects two nodes of degrees $k$ and $k'$, $q(k)$ is the probability that a randomly chosen link is attached to a node with degree $k$
\begin{equation}
q(k)=\frac{kp(k)}{\langle k \rangle},
\end{equation}
$p(k)$ is the degree distribution of the network, i.e., the probability that a node chosen uniformly at random from the network has degree $k$, and $\langle k \rangle = \sum_k kp(k)$ is the average degree over the whole network. 

In uncorrelated networks, the joint probability $p(k, k')$ factorizes and can be expressed in terms of the degree distribution, i.e., $p(k,k')=\frac{kk'}{\langle{k} \rangle ^2}p(k)p(k')$, thus yielding

\begin{equation}
\label{k2}
K_{nn}(k) =  \sum_{k'}k'\frac{p(k, k')}{q(k)}=\frac{\langle k^{2} \rangle}{\langle k \rangle}.
\end{equation}

Thus, if there are no degree correlations, $K_{nn}(k)$ does not vary as a function of $k$: regardless of the degree a node has, its nearest neighbors have on average the same degree. By contrast, an increasing (decreasing) behavior of $K_{nn}(k)$ as a function of $k$ indicates that the network is assortative (disassortative) by degree: as the degree of a node increases, the degree of the node's nearest neighbors tends, on average, to increase (decrease). 

The second method for detecting degree correlations relies upon the assortativity coefficient, a measure originally proposed by Newman \cite{Newman2002} that is a suitably modified version of the standard Pearson correlation coefficient for measuring the correlation between the degrees of adjacent nodes in a network. Given a randomly chosen node that lies at the end of a randomly chosen link, one can define the excess degree of that node as the number of links incident upon the node other than the one along which the node was reached \cite{Newman2001}. The excess degree of such node is distributed according to
\begin{equation}
e(k)=\frac{(k+1)p(k+1)}{\langle k \rangle}.
\end{equation}

The assortativity coefficient for detecting mixing by degree can now be defined as 
\begin{equation}
\label{eq1}
r=\frac{1}{\sigma_e^2}\sum_{kk'}kk'(e(k, k')-e(k)e(k')),
\end{equation}
where $e(k, k')$ is the joint probability that a randomly chosen link in the network connects a node that has excess degree $k$ with a node with excess degree $k'$, $\sigma_e^2=\sum_k k^2e(k) - \left [\sum_k k e(k)\right]^2$ is the variance of the distribution $e(k)$, and $e(k)e(k')$ is the expected value of the quantity $e(k, k')$ in the case in which links are placed between nodes uniformly at random regardless of the degrees of the connected nodes. The values of $r$ lie in the range $-1 \le 0 \le 1$, with $r=1$ indicating perfect assortativity, $r=-1$ perfect disassortativity, and $r=0$ lack of degree correlations \cite{Newman2002}.

\begin{figure}
\centering
\includegraphics[scale=0.55]{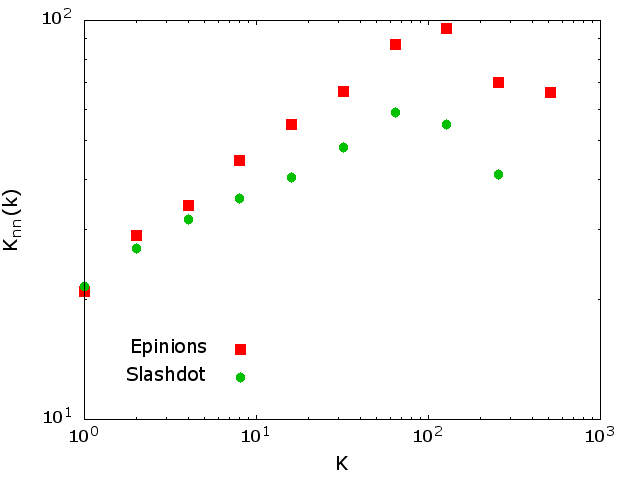} 
\caption{\textbf{$K_{nn}(k)$ for the Slashdot and Epinions unsigned networks with reciprocated links}. The observed positive trends are in qualitative agreement with the assortative patterns found in many other social networks. Data were logarithmically binned.}
\label{Knn_unsigned}
\end{figure}

\begin{figure} 
\centering
\includegraphics[scale=0.55]{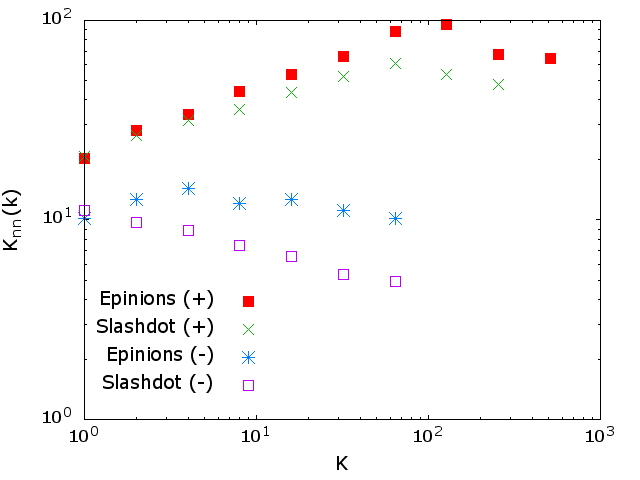}  
\caption{\textbf{$K_{nn}(k)$ for the Epinions and Slashdot positive and negative subnetworks}. The positive subnetworks display a positive trend, while the negative subnetworks display a negative trend. Data were logarithmically binned.}
\label{knn_super}
\end{figure}

We begin our analysis of degree correlations by uncovering mixing patterns from the unsigned Epinions and Slashdot networks with reciprocated links. Fig.~\ref{Knn_unsigned} shows a positive trend for $K_{nn}(k)$, as was typically documented in social networks. To analyze degree correlations in the signed subnetworks, we measured and plotted $K_{nn}(k)$ for all four subnetworks. As shown in Fig.~\ref{knn_super}, two main distinct patterns can be detected. The positive subnetworks show the typical structural signature of social networks, namely the tendency of nodes to connect to other nodes with a similar degree (assortative mixing by degree). By contrast, the negative subnetworks display disassortative mixing by degree: high-degree nodes tend to be connected with low-degree ones. 

This finding is further corroborated by the values obtained for the correlation coefficient $r$. These values are $r_{Ep}^+=0.217$ and $r_{Ep}^-=-0.022$ for the positive and negative Epinions subnetworks, respectively, and $r_{Sl}^+= 0.162$ and $r_{Sl}^-=-0.114$ for the positive and negative Slashdot subnetworks, respectively. All coefficients are statistically significant (\textit{p-values} $<0.0001$), with the only exception of $r_{Ep}^-$ ($p>0.05$).

These results are in qualitative agreement with, and generalize, a widely supported empirical regularity found in a variety of social networks: when links have a positive connotation, or can be assumed to have a positive one, they tend to connect nodes with similar degrees \cite{Newman2002, Newman2003}. However, our findings also suggest that, when links have a negative connotation, they tend to connect nodes with dissimilar degrees \cite{Szell2010, Szell2010a}. Combined, these two sets of results undercut one of the arguments that the literature has proposed to explain degree correlations in social networks \cite{Newman2003}. This argument is premised on the idea that assortative mixing is attributable to the tendency of nodes to coalesce into distinct communities. However, because this tendency can be detected in both the positive and negative subnetworks, community structure would in itself be not sufficient for explaining the assortative mixing patterns observed in the positive subnetworks. Other mechanisms are likely to be responsible for these patterns. 

In both Epinions and Slashdot, individuals cluster into communities based on their common interests in the same products or news. However, the observed mixing patterns seem to originate not simply from common interests, but more precisely from the way individuals use the posted comments as cues for making positive or negative judgements on one another. More generally, the comparison between positive and negative subnetworks suggests that the observed degree correlations depend on the sign of the links between nodes \cite{Szell2010a}. To gain a better understanding of this relation between sign of links and degree correlations, in what follows we shall propose a class of simple generative models of signed networks.

\section{Signed networks with degree correlations that depend on the sign of the links}
\label{model}
We begin by focusing on signed random networks with binomial degree distributions, in which nodes can be split into two mutually exclusive groups. Subsequently, we shall refine our analysis by investigating the case of assortative and disassortative signed networks with power-law degree distributions. 

\subsection{Signed random networks with binomial degree distributions}
\label{random}

We draw on, and extend, a model originally developed by Newman and Park for undirected unsigned networks with multiple communities \cite{Newman2003}. We create random networks with $N$ nodes that satisfy the following requirements: 

\begin{enumerate}
\item degrees are homogeneously distributed across the nodes; 
\item each node can be a member of one of two mutually exclusive groups;
\item there are no degree correlations prior to the attribution of signs to the links; and 
\item signs are associated with links in such a way that the resulting signed network is structurally balanced, i.e., it contains only positive cycles ~\cite{Cartwright1956, HEIDER1946a}. 
\end{enumerate}

To obtain such networks, we apply the following rules: 

\begin{enumerate}

\item any pair of nodes are connected through a link with a uniform probability $p$;
\item given two groups, each node is assigned to one of them with probability $m$ and to the other with probability $1-m$; and
\item connections between nodes within the same group are associated with a positive sign, while connections between nodes from different groups with a negative sign. 
\end{enumerate}

\begin{figure}
\centering
\includegraphics[scale=0.25]{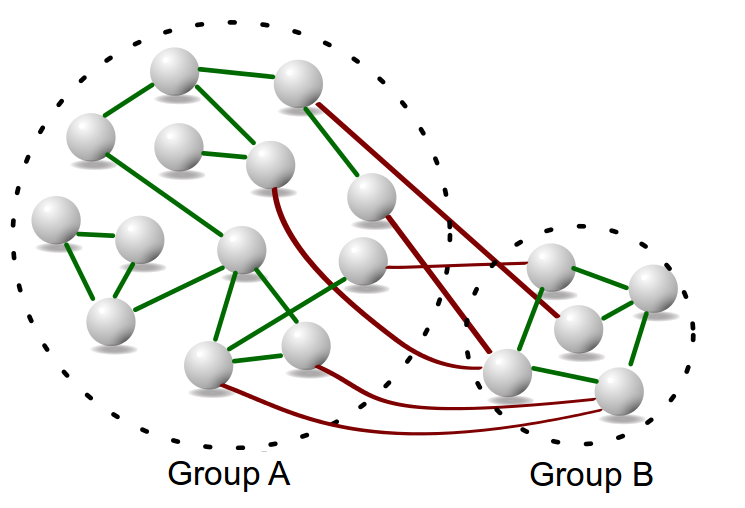}
\caption{\textbf{Network polarization and sign attribution}. Schematic representation of the allocation of nodes into two mutually exclusive groups. Links between nodes belonging to the same group are positive (green), whereas links between nodes of different groups are negative (red).}
\label{two_groups_division}
\end{figure}

A schematic representation of the polarization of a network into two distinct groups according to our model is shown in Fig.\ref{two_groups_division}. The model generates random uncorrelated networks with a binomial degree distribution. Notice that the original model proposed by Newman and Park corresponds to the case in which there is more than one community and $m=1$ such that the resulting network is unsigned by construction. In our case, for the sake of simplicity, we introduced only one community as findings are qualitatively similar to those obtained with multiple communities. Moreover, as $m$ approaches the value of $0.5$, the network becomes perfectly polarized into two distinct groups of equal size. As $m$ gets closer to either zero or one, polarization gradually disappears, and the network becomes increasingly dominated by one of the two groups \cite{Blau1977}. 

Finally, to obtain a signed network, we attribute signs to links using an assignment rule that discriminates between links within and across groups. According to the \textit{structure theorem} \cite{Cartwright1956, Harary53}, any signed network polarized into two mutually exclusive subsets of nodes, such that each positive link connects two nodes of the same subset and each negative link connects nodes from different subsets, will include an even number of negative links. In accordance with the definition originally proposed by Heider \cite{HEIDER1946a} and subsequently extended by Cartwright and Harary \cite{Cartwright1956}, this is indeed the signature of structural balance \cite{Altafini2013}. Thus, the application of our rule of sign attribution will ensure the generation of structurally balanced networks.
 
As with the real networks, from the global signed network we extract two subnetworks, each including only positive or negative links. We then test whether and the extent to which network polarization has any critical role in the emergence of non-trivial mixing patterns in the positive and negative subnetworks. To this end, we simulate the model for an arbitrarily large value of $N$, and calculate the values of the correlation coefficient $r$ between the degrees of connected nodes within the unsigned networks and the signed subnetworks obtained in correspondence of the different values of the probability $m$. 

\begin{figure}
\centering
\includegraphics[scale=0.4]{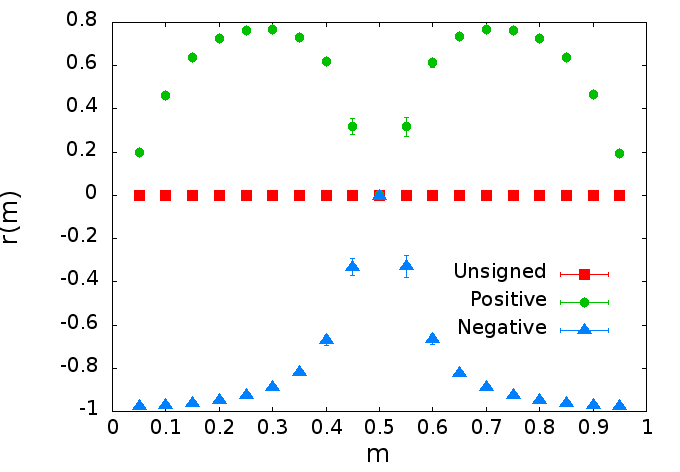}
\caption{\textbf{Correlation coefficient $r$ plotted against the probability $m$ of being a member of one group}. The graph shows the trends of $r$ for the positive and negative subnetworks obtained with the model, when $N=10,000$ and $p=0.01$. For each value of $m$,  the correlation coefficient $r$ is the average over $50$ realizations of the network.}
\label{r_vs_mainstream}
\end{figure}

As indicated by Fig.~\ref{r_vs_mainstream}, the positive subnetwork displays an assortative mixing by degree, as was observed in our two positive subnetworks as well as in many other social networks documented in the literature \cite{Newman2002,Newman2003}. By contrast, the negative subnetwork, like the ones extracted from both the Epinions and Slashdot networks, displays a disassortative mixing pattern. The sign of the links or, more precisely, the rules underpinning the attribution of sign to links, seem to be responsible for the variation in the mixing patterns. In particular, results suggest that non-trivial degree correlations of the signed networks would remain hidden and undetected if they were simply assumed to be the same as the ones of the corresponding unsigned networks obtained by removing or ignoring the signs of the links. 

To further explore the conditions under which such degree correlations are likely to emerge in signed networks, in the subsequent section we shall extend our analysis by using a number of more refined and realistic network generative models and by introducing additional combinations of structural properties of the networks. However, before we proceed in that direction, we now formalize the properties of our current model in terms of the degree correlations displayed by the signed subnetworks. Given $N$ nodes, and two mutually exclusive groups $A$ and $B$, we set $N_A$ to be the number of nodes that belong to group $A$, and $N_B = N - N_A$ the number of nodes that belong to group $B$. The probability that in group $A$ there are $N_A$ nodes can be expressed as

\begin{equation}
p(N_A)={N \choose N_A} m^{N_A}(1-m)^{N-N_A}.
\label{pna}
\end{equation}
\noindent 
The ``positive" degree $k_{A,+}$ of a node in group $A$ is the number of positive links incident upon the node. The probability that a node that belongs to group $A$ has a ``positive" degree  $k_{A,+}$  when the total number of nodes in group $A$ is $N_A$, is given by 

\begin{equation}
\label{pkp}
p(k_{A,+}|N_A) = {N_A-1 \choose k_{A,+}} p^{k_{A,+}}(1-p)^{N_A-k_{A,+}}.
\end{equation}

\noindent  In Eq.\ref{pkp}, $p$ represents the independent probability of a link in the network, and $k_{A,+}$ is the positive degree of nodes in group A (i.e., the number of links to other nodes in $A$). We define  $K^{A,+}_{nn}(k_{A,+})$ as  the average positive degree of the nearest friends of nodes of group $A$. Since, given a certain number of nodes in group $A$, the network formed by the links between these nodes is a random network (i.e., uncorrelated), using Eq.\ref{k2} we have 
\begin{equation}
\label{knn}
K_{nn}^{A, +}(k_{A,+}) =\sum_{N_A>0}p(N_A) \frac{\langle(k_{A,+})^2|N_A\rangle}{\langle{k_{A,+}|N_A}\rangle}\simeq Npm,
\end{equation}
where the average in Eq.\ref{knn} is taken over the distribution $p(k_{A,+}|N_A)$ defined in Eq.\ref{pkp}, $P(N_A)$ is defined in Eq.\ref{pna}, and where we have assumed $N\gg 1$.

We thus obtained a constant value for $K_{nn}^{A,+}(k_{A,+})$ that is independent of the positive degree $k_{A,+}$. In the same way, if we evaluate $K_{nn}^{B,+}(k_{B,+})$, i.e., the average positive degree of the nearest friends of nodes in group $B$, we obtain: $K_{nn}^{B,+}(k_{B,+})=Np(1-m)$, which is also a constant function of $k_{B,+}$. As to the negative subnetwork, we obtain the same results for both groups of nodes. That is, $K_{nn}^{A,-}(k_{A,-}) = Npm$ is the average negative degree of the nearest enemies of nodes in  group $A$, and $K_{nn}^{B,-}(k_{B,-})=Np(1-m)$ is the average negative degree of the nearest enemies of nodes in group $B$. What differentiates the two groups in each signed subnetwork is simply the mean degree of their nodes. For instance, if nodes of group $A$ (or $B$) in the positive subnetwork have an average positive degree of $Npm$ (or $Np(1-m)$), in the negative subnetwork they have an average negative degree of $Np(1-m)$ (or $Npm$). In the case of $m=0.5$, namely when groups are of equal size, it would not be possible to distinguish between the two subnetworks (see Fig.\ref{r_vs_mainstream}), and we thus obtain the same results as in the case of the uncorrelated unsigned network.

\begin{figure}
\centering
\includegraphics[scale=0.3]{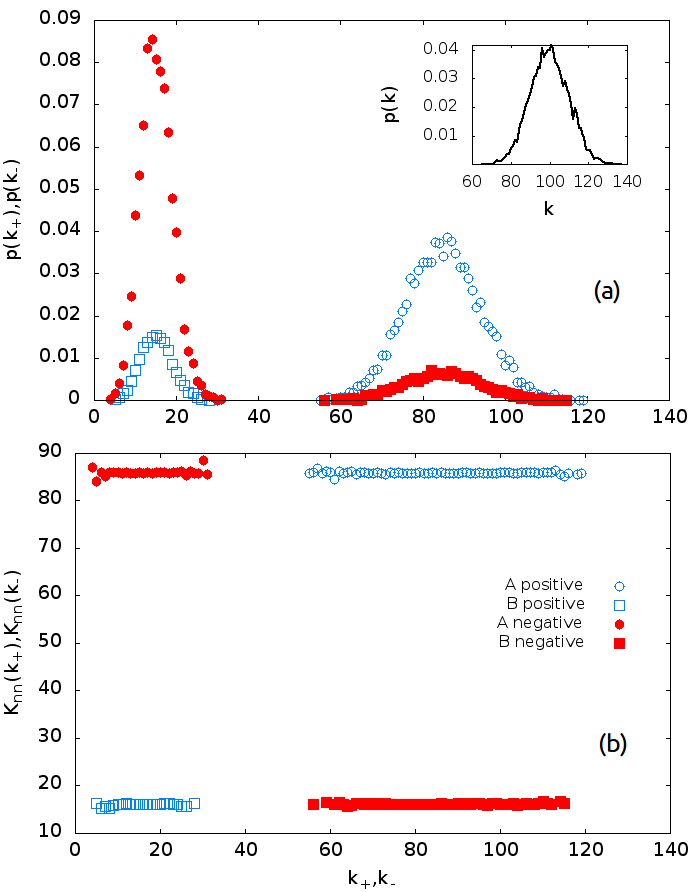}
\caption{\textbf{The positive and negative degree distributions $p(k_+)$ and $p(k_-)$ and the trends of $K^+_{nn}(k_+)$ and $K^-_{nn}(k_-)$ for a network with a binomial unsigned degree distribution and polarization into two groups}. A network with a binomial degree distribution was created, with $N=10,000$, $p=0.01$ and $m=0.85$, and in which $A$ is the larger group. Panel (a) shows the positive and negative degree distributions, $p(k_+)$ and $p(k_-)$. Findings indicate the two distinct distributions for each signed subnetwork, one attributable to group $A$ and the other to group $B$. The inset shows the degree distribution of the unsigned network. Panel (b) displays the trend of $K_{nn}^+(k_+)$ and $K_{nn}^-(k_-)$, respectively for the positive and negative subnetworks. For each subnetwork, the value of $K_{nn}(k)$ is constant within the same group, i.e., $85$ for group $A$ and $15$ for group $B$. The two panels indicate that there are two corresponding gaps between values for $K_{nn}(k)$ and the signed degree distributions. This is due to the fact that the minimum value of degree in the unsigned network is $60$ (see inset). Each node, regardless of the group it belongs to, has on average $85\%$ of its neighbors from group $A$. If the node with degree $60$ belongs to group $A$ ($B$), it has, on average, $k_+=55$ ($k_-= 5$) in the positive (negative) subnetwork. Similarly, the maximum negative (positive) degree for a node in group $A$ ($B$) would depend on the maximum value of the degree in the unsigned network, i.e., $140$, yielding $k_-=28$ ($k_+=119$). This therefore causes a gap between degrees ranging from $28$ to $55$, as shown in both panels. Panel (b) indicates the positive and negative trends for $K_{nn}(k)$, respectively for the positive and negative subnetworks, when both contributions from the two groups are taken into account.}
\label{knn_Poissoniano}
\end{figure}

As suggested by Eq.\ref{knn}, the polarization of a network with a binomial degree distribution into two groups of heterogeneous size generates two distinct values of $K_{nn}(k)$ for each of the two subnetworks, namely $Nmp$ and $(1-m)Np$. In other words, the overall values of $K_{nn}(k)$ for each signed subnetwork result from the different (and complementary) contributions of the two groups in which, in turn, nodes have positive and negative degrees of different (and non-overlapping) values (see Fig.\ref{knn_Poissoniano}a). For instance, in the case of the positive subnetwork, and when $m > 0.5$ and $A$ is the larger group, the contribution to the overall $K_{nn}^+(k_+)$ from group $A$ is $K_{nn}^+(k_{(A,+)})=Npm$ (which in turn corresponds to the larger values of $k_+$), while the contribution from group $B$ is $K_{nn}^+(k_{(B,+)})=Np(1-m)$ (which corresponds to the smaller values of $k_+$). As indicated by Fig.\ref{knn_Poissoniano}b, when the two contributions are combined, $K_{nn}^{+}(k_+)$ takes on two distinct constant values in correspondence of two distinct sets of values of the positive degree, thus yielding the positive trend that signals the assortative mixing pattern of the positive subnetwork. Similarly, the negative trend of $K_{nn}^{-}(k_-)$ for the disassortative negative subnetwork results from the combination of the two distinct and complementary contributions from the two groups: $K_{nn}^-(k_{(A,-)})=Npm$ from group $A$ in correspondence of the smaller values of $k_-$, and $K_{nn}^-(k_{(B,-)})=Np(1-m)$ from group $B$ in correspondence of the larger values of $k_-$. 

These trends are primarily due to the value of $m$ which, in turn, affects the \textit{opportunity} for nodes to create links within and across groups. Notice that, on average, the value of the degree of a randomly chosen node from a network with a standard binomial degree distribution is $Np$, regardless of which group the node belongs to. However, the polarization of the network into two groups of unequal size (i.e., $m \neq 0.5$), in combination with our rule of sign attribution, generates heterogeneity across nodes in terms of the proportion between positive and negative links incident upon them. Let us suppose that group $A$ is the larger one. Each node, regardless of the group it belongs to, is surrounded approximately (for large $N$) by $Nm$ potential neighbors from the dominant group ($A$) and $(1-m)N$ potential neighbors from the smaller group ($B$). Thus, each node, regardless of its affiliation, is likely to direct most of its links toward the nodes that belong to the larger group. This, in turn, has a direct bearing on the relative number of friends and enemies a node can have, depending on the group it belongs to. Because a node that belongs to the larger group has a higher chance than a node from the smaller group to direct links toward nodes of its own group (i.e., $A$), then as a result of our rule of sign attribution a node from the larger group also has a higher chance than a node form the smaller group to create friends by forging positive links with others. By contrast, a node from the smaller group ($B$) is more likely than a node from the larger group ($A$) to create links across groups, which in turn leads the former node also to be more likely to create more negative links than the latter. This difference in opportunity of ``signed interactions" is responsible for the two different values obtained for the positive and negative $K_{nn}(k)$ attributable to the two groups of nodes, and can ultimately explain the assortative and disassortative mixing patterns, respectively of the positive and negative subnetworks.

\subsection{Signed networks with power-law degree distributions}
\label{power}
Previous empirical research has documented a large number of social networks characterized by statistically heterogenous connectivity: while the majority of nodes have only few connections, a minority have a disproportionally large amount of links to other nodes \cite{Barabasi1999}. For this reason, we now move beyond the case of random networks with binomial degree distributions, and study the mixing patterns of more realistic signed networks characterized by power-law degree distributions. To this end, we introduce a generative model of scale-free signed networks. The choice of the model is also motivated by the need to ensure that the resulting unsigned network (i.e., the network obtained prior to the allocation of signs to links) is characterized by non-trivial degree correlations. This, in turn, will serve a two-fold purpose. First, it will help create networks with structural properties that are comparable to those observed in a variety of real-world networks \cite{Newman2002, Pastor-Satorras2001, Park2003}. Second, it will allow us to investigate whether the degree correlations of the unsigned network may be responsible for the difference between the mixing patterns of the positive and negative subnetworks.

We begin by constructing unsigned networks characterized by a power-law distribution and assortative mixing by degree. This will enable us to replicate the patterns observed in both the Slashdot and Epinions unsigned networks (see Fig.\ref{Knn_unsigned}). Among the models that satisfy the above requirements, in what follows we shall use the copying model \cite{Kumar1999} and an extension of the rewiring model proposed by Xulvi-Brunet and Sokolov \cite{Xulvi2004} based on the scale-free Barab\'{a}si-Albert network \cite{Barabasi1999}. 

First, the copying model begins with an initial connected network of $n$ nodes. At each step, a new node is added to the network and another incumbent node is selected by chance: with a probability $p$ the new node will create a link with one of the neighbors of the selected node, and with a probability $1-p$ it will create a link with a node selected at random. Second, the rewiring model \cite{Xulvi2004} is suitably applied to an initial network with a given scale-free degree distribution obtained by following the rules of the Barab\'{a}si Albert model \cite{Barabasi1999}. The rewiring process is then modeled as follows: (i) two links are selected at random; (ii) the four nodes connected through these two links are sorted in increasing order of degree; (iii) if the first two nodes and the last two nodes are not connected, links are rewired accordingly; otherwise, (iv) the two links are dismissed, and a new pair of links are selected. After several iterations, an assortative network can be obtained. Both methods indeed generate an unsigned, undirected and assortative network characterized by a power-law degree distribution.

Drawing on these two generative models, we obtain unsigned networks that we then transform into signed networks by applying the last two rules from the basic model in Section \ref{model}, namely: (i) polarization of the network into two mutually exclusive groups of nodes; and (ii) attribution of a positive sign to links within groups and a negative sign to links across groups. Just as with the uncorrelated case, we then extract the positive and negative subnetworks from the signed networks, detect the mixing patterns of these subnetworks, and compare them with the patterns observed in the unsigned network. 

To shed light on the role of the sign of links in the emergence of mixing patterns, we vary the rules governing network polarization and sign attribution, and extract and assess the corresponding signed subnetworks. First, as with the uncorrelated case, we manipulate network polarization by varying the degree to which the two groups differ in size. To this end, we use different values of $m$, the probability that a node belongs to one of two groups: as usual, at $m=0.5$, the network is perfectly polarized, while for values approaching zero and one, the network becomes increasingly homogeneous and dominated by one single group \cite{Blau1977}.  

Second, by manipulating our rule of sign attribution, we aim to vary the degree to which the signed network is structurally balanced. Previous research has long provided empirical evidence in favor of the tendency of individuals to avoid or alleviate cognitive tension by transforming an unbalanced structure into a balanced one \cite{Doreian1996,Facchetti2011}. Yet, a number of studies have equally suggested that many observed signed structures for social groups are not structurally balanced, at least when they are assessed at single points in time \cite{Doreian2001, Doreian2009}. To account for such empirically documented variations in structural balance, in what follows we test whether this property, in combination with other conditions, is indeed necessary for the emergence of non-trivial mixing patterns within signed networks that differ from those observed in the corresponding unsigned networks. In this sense, we extend our previous analysis by investigating whether the sign of links can still produce some effect upon degree correlations also when the network is unbalanced. 

Notice that, as implied by the structure theorem \cite{Cartwright1956, Harary53}, to obtain structurally unbalanced networks, it would not be possible to divide the population of nodes into two even or uneven groups and then impose our homophily-based rule of sign attribution (i.e, positive links within groups and negative links across groups). Indeed, from the structure theorem it follows that this procedure would necessarily generate a structurally balanced network. To obtain an unbalanced network, we therefore reshuffle the signs of the links within the corresponding balanced networks. In this way, the random reallocation of signs to links transforms the network from a balanced to an unbalanced state. 

In summary, starting from assortative unsigned networks with a power-law degree distribution, we create four distinct groups of signed networks and corresponding subnetworks by combining the following structural conditions: (i) even versus uneven allocation of nodes into two mutually exclusive groups; and (ii) balanced versus unbalanced network structure. For the sake of simplicity, we label the four groups of networks as follows: 

\begin{enumerate}
\item[] \textbf{Ass/Het/Bal}: (i) The unsigned network is assortative; (ii) nodes are heterogeneously allocated to groups; and (iii) the signed network is balanced. 
\item[] \textbf{Ass/Hom/Bal}: (i) The unsigned network is assortative; (ii) nodes are homogeneously allocated to groups; and (iii) the signed network is balanced. 
\item[] \textbf{Ass/Het/Un}: (i) The unsigned network is assortative; (ii) nodes are heterogeneously allocated to groups; and (iii) the signed network is unbalanced. 
\item[] \textbf{Ass/Hom/Un}: (i) The unsigned network is assortative; (ii) nodes are homogeneously allocated to groups; and (iii) the signed network is unbalanced. 

\end{enumerate}

\begin{figure}
\centering
\includegraphics[scale=0.25]{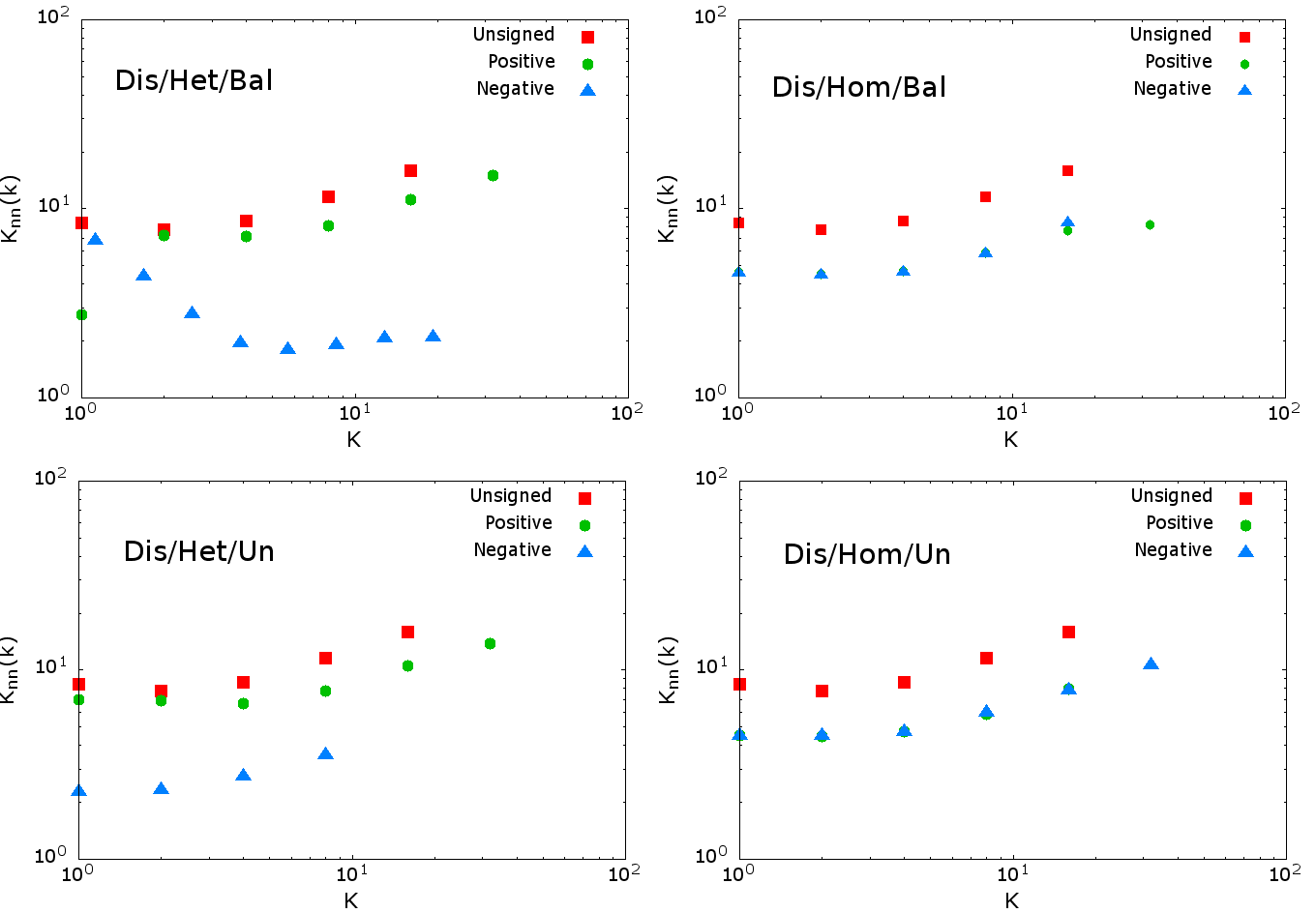}
\caption{\textbf{Positive and negative subnetworks obtained from an assortative unsigned network with power-law degree distribution}. The unsigned network was generated through the copying model with $N=10^4$ nodes. Findings indicate that different mixing patterns for the positive and negative subnetworks are obtained only when the assortativity of the unsigned network is combined with the heterogeneous allocation of nodes into groups and with the presence of structural balance. In all panels, data were logarithmically binned.}
\label{ass_net}
\end{figure}

Results are shown by Fig.\ref{ass_net}, in which the unsigned assortative networks were generated through the copying model \cite{Kumar1999}. Each panel of Fig.\ref{ass_net} shows the trends of $K_{nn}(k)$ for the unsigned network, for the positive subnetwork, and for the negative subnetwork obtained under each of the four combinations of structural conditions. Findings clearly indicate that most signed subnetworks retain the assortative pattern that characterizes their corresponding unsigned networks. There is, however, an exception: as indicated by panel (a) of Fig.\ref{ass_net}, there is one case in which a decreasing trend of $K_{nn}(k)$ for the negative subnetwork is associated with an increasing trend for the unsigned network and the positive subnetwork. In particular, this opposite trend in mixing patters occurs when the following three conditions are jointly satisfied:  
  
\begin{enumerate}
\item the unsigned network is assortative;
\item nodes are unevenly allocated into two mutually exclusive groups; and
\item the signed network is structurally balanced.
\end{enumerate}

Under the above conditions, the disassortative pattern of the negative subnetwork would therefore remain hidden if the signs of links were removed from the global signed network and the nature and intensity of the mixing patters were simply inferred from the resulting unsigned network. Similar results are obtained when the assortative unsigned network is created by applying the rewiring model by Xulvi-Brunet and Sokolov \cite{Xulvi2004} to the scale-free Barab\'{a}si-Albert network \cite{Barabasi1999}. In this case, once again the negative subnetwork exhibits a variation in mixing patterns and becomes disassortative when the unsigned network is assortative, the groups are uneven in size, and the signed network is balanced. 

\begin{figure}
\centering
\includegraphics[scale=0.25]{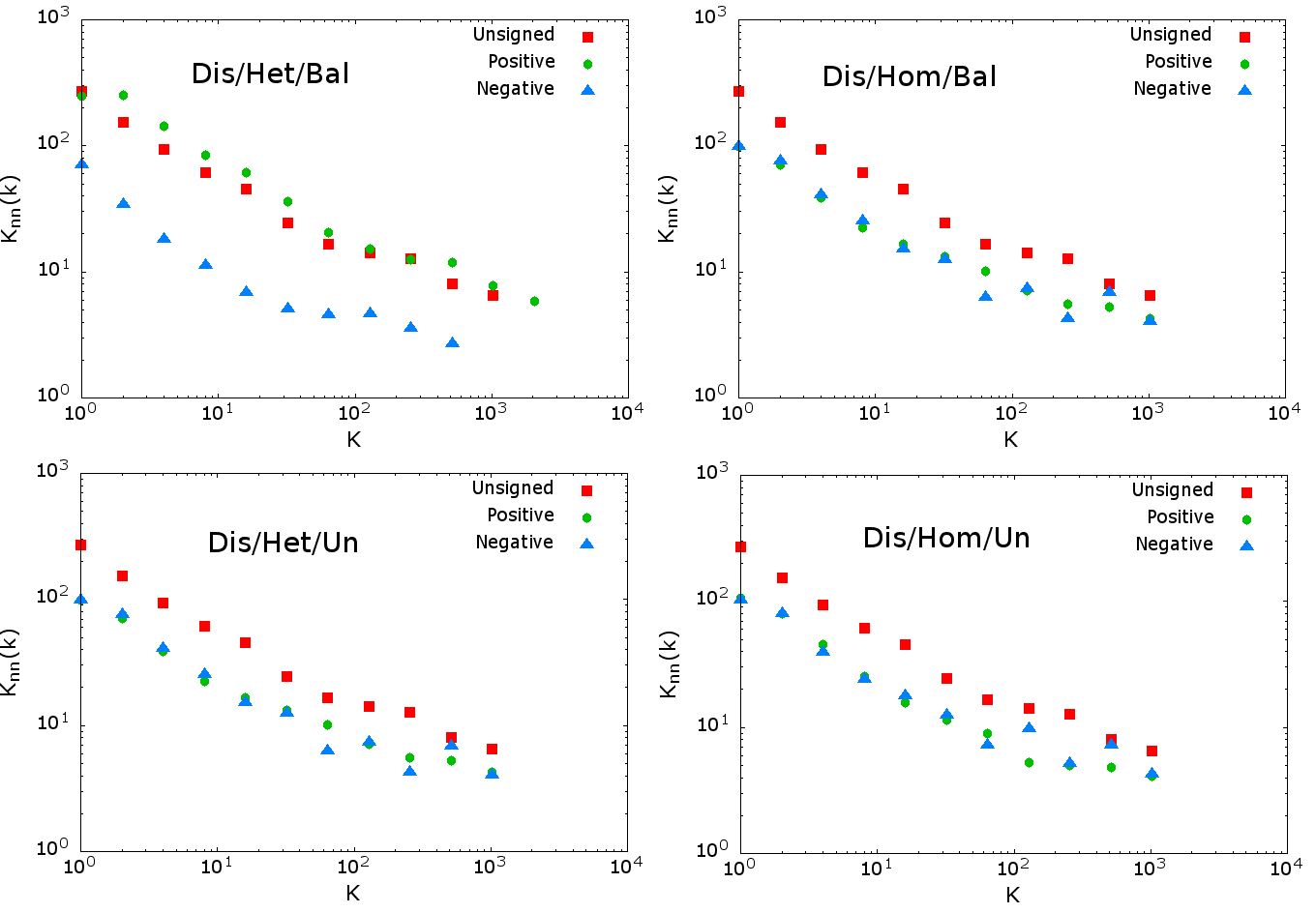}
\caption{\textbf{Positive and negative subnetworks obtained from a disassortative unsigned network}. The unsigned network was obtained by removing the signs from the links of the network generated through the fitness model of growing networks proposed by Bianconi and Barab\'{a}si \cite{Bianconi2001}, with $N=10^4$ nodes. The unsigned network is characterized by a power-law degree distribution. Results indicate that across all combinations of the three conditions the mixing patterns for the positive and negative subnetworks have the same trend. In all panels, data were logarithmically binned.}
\label{bara_net}
\end{figure}

We now test whether the mixing patterns in the positive and negative subnetworks differ when the unsigned network is disassortative. To this end, we create an unsigned network following the rules of the fitness model of growing networks, originally proposed by Bianconi and Barab\'{a}si \cite{Bianconi2001}. The results from our simulations are shown by Fig.\ref{bara_net}, in which the trend of $K_{nn}(k)$ is reported. If the unsigned network is characterized by a disassortative pattern, the patterns for the positive and negative subnetworks will always have the same trend across any of the four possible combinations of our two initial conditions. Subnetworks will always retain their disassortativity, regardless of the structural balance of the global network and the size of the groups. 

Table \ref{tab_r} reports the correlation coefficient $r$ of the degrees of connected nodes, for each of the networks and subnetworks analyzed above. The Table clearly indicates that there is only one case in which the mixing patterns of the positive and negative subnetworks differ. This variation indeed occurs when the unsigned network is assortative, the signed one is balanced, and groups differ in size. Under this combination of structural conditions, the correlation coefficient becomes negative for the negative subnetwork, while it remains positive for the positive one. Similar results are obtained when the assortative unsigned network is created by using the rewiring model by Xulvi-Brunet and Sokolov \cite{Xulvi2004}.

\begin{table}
\centering
\begin{tabular}{|l|c|c|}
\hline
\multicolumn{3}{ |c| }{\textbf{The case of two groups}} \\
\hline 
	\textbf{Conditions}&\textbf{Dis. Unsigned Network} &\textbf{Ass. Unsigned Network}\\
\hline
					&$r^u=-0.09413$                & $r^u=0.16249$     \\
\textbf{Het/Bal}    &$r^+=-0.09853$                & $r^+=0.15598$     \\
                &$r^-=-0.10864$                & $r^-=-0.3509$     \\
                \hline
                &$r^u=-0.09413$                & $r^u=0.16249$     \\
\textbf{Hom/Bal}    &$r^+=-0.09199$                & $r^+=0.12062$     \\
                &$r^-=-0.09692$                & $r^-=0.14804$     \\
                \hline
                &$r^u=-0.09413$                & $r^u=0.16249$     \\
\textbf{Het/Un}    &$r^+=-0.093570$                & $r^+=0.15243$     \\
                &$r^-=-0.096497$                & $r^-=0.08244$     \\
                \hline
                &$r^u=-0.09413$                & $r^u=0.16249$     \\
\textbf{Hom/Un}    &$r^+=-0.09748$                & $r^+=0.12596$     \\
                &$r^-=-0.090807$                & $r^-=0.13250$     \\      						
\hline 
\end{tabular}
\caption{\textbf{Values of the correlation coefficient $r$ for the case of polarization of the network into two groups.} The coefficient was calculated for each of the four combinations of structural balance (Bal) and unbalance (Un), and even (Hom) and uneven (Het) group size. Under each of the four combinations, the coefficient was calculated distinctively for each of the two cases in which the unsigned global network is assortative (and obtained through the copying model) and disassortative (and obtained with the fitness model). The variation in sign of the correlation coefficient between the positive and negative subnetworks occurs only when the unsigned network is assortative, the signed network is balanced, and groups are of unequal size.}
\label{tab_r}
\end{table}

The reason for the opposite trends  in the mixing patterns of the two signed subnetworks is similar to the one that explains the transformation of an unsigned uncorrelated random network into correlated signed subnetworks. As before, this reason is two-fold. First, the polarization of the network into two groups of unequal size is responsible for the heterogeneous distribution across nodes of opportunities of creating links within and across groups. Second, the requirement of structural balance (i.e., the rule of sign attribution) transforms these heterogeneous opportunities of social contact into equally heterogeneous opportunities to create friends or enemies. While a node of the larger group has a higher chance than a node of the smaller group to create intra-group connections, the latter node will have a higher chance to create inter-group connections than the latter. This imbalance of opportunities will be translated into the differential propensity nodes will have to create friends or enemies, depending on which group they belong to. It then follows that, when the whole unsigned network is assortative (disassortative), the positive subnetwork will remain assortative (disassortative) as it only includes intra-group connections between nodes of comparable propensity to make friends. Conversely, because the negative subnetwork only includes inter-group links, it will connect nodes that differ in their propensity to make enemies. For this reason, it will always remain disassortative, also when the unsigned network is assortative.  

\section{Extending the model: The case of three (or more) groups}
\label{three}
 
Following the theoretical avenue that led Davis \cite{Davis1967} to generalize the formalization of the theory of structural balance, we extend our model with network polarization to also account for the case in which nodes can be allocated to three or more mutually exclusive groups. As observed by Davis \cite{Davis1967}, individuals often split into more than two mutually hostile groups. To take this into account, Davis provided a generalization of the structure theorem \cite{Cartwright1956, Harary53} by uncovering the necessary and sufficient condition for a signed network to be clusterable into two or more groups of nodes such that links connecting nodes within the same group are positive, and links connecting nodes from different groups are negative. Such condition was identified in the absence of cycles with exactly one negative link. It follows that all structurally balanced networks are clusterable, but not vice versa. Whether clusterable networks are also balanced depends on the number of disjoint groups of nodes. 

The analysis carried out by Davis provides us with a theoretical backdrop against which we can further refine our model. First, we investigate whether our model is robust against the number of groups, namely whether the same results are obtained when the network splits into more than two mutually exclusive groups, but still remains structurally balanced. Second, we study our model in the more general case in which the network is clusterable into more than two groups, but it is not balanced. In what follows, we shall focus our attention on the case of three groups, and then briefly outline how the analysis can be generalized to any number of mutually exclusive groups.  

\begin{figure}
\centering
\includegraphics[scale=0.25]{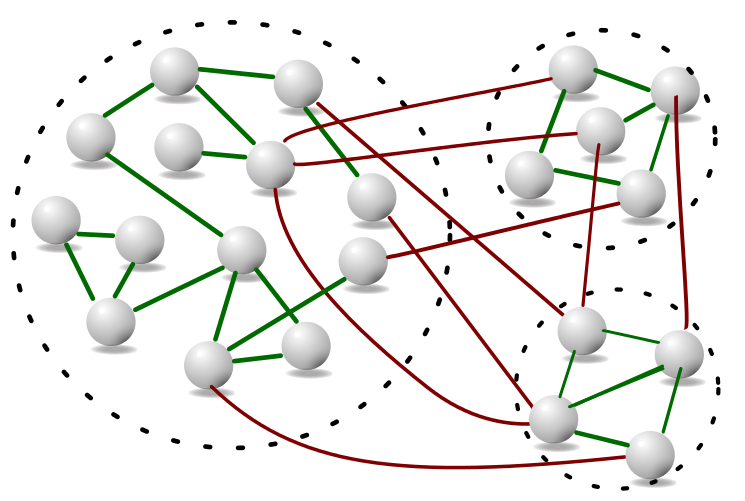}
\caption{\textbf{The case of three mutually exclusive groups}. Schematic representation of the allocation of nodes into three groups such that links connecting nodes of the same group are positive (green), and links between nodes from different groups are negative (red).}
\label{three_groups}
\end{figure}

Fig.\ref{three_groups} shows a schematic representation of a network that splits into three mutually exclusive groups. The rule of sign allocation remains the same as before: links between nodes of the same group are assumed to be positive, and links between nodes from different groups negative. Let us assume that each node can belong to one of the three groups with a given probability $p$. We then have four possible cases:

\begin{enumerate}
\item[1.] homogeneous allocation of nodes into groups of equal size, i.e., $p_1=p_2=p_3$;
\item[2.] heterogeneous allocation of nodes into groups of uneven size, such that one group dominates the other two, i.e., $p_1 > p_2 \simeq p_3$;
\item[3.] heterogeneous allocation of nodes into groups of uneven size, such that two equally sized groups dominate a less populated one, i.e., $p_1 \simeq p_2  > p_3$; and
\item[4.] heterogeneous allocation of nodes into groups of uneven size, such that, for any two groups, one dominates the other, i.e., $p_1 > p_2 > p_3$.
\end{enumerate}

In what follows, we shall concentrate on the first two cases. Results concerned with the third case will not be reported here because they are qualitatively similar to what is obtained with: (i) two equally sized groups, when the two dominant groups are much larger than the third one; and (ii) three equally sized groups, when differences in size become negligible. Similarly, the fourth case can be reduced to the previous cases, depending on the difference in size between groups. 

To create a structurally balanced network, we impose the following constraint. When there is a (negative) link between two nodes that belong to two different groups, the two connected nodes are not allowed to share a common enemy, that is they are not allowed to be connected with the same node from the third group. In this case, each of the two nodes will change the target of the link to the third group, so as to avoid triangles in which all links are negative. In other words, two nodes may share a common enemy either when they are not connected themselves, or when they are connected and belong to the same group. In this sense, allowing coalition formation against a common enemy to occur only between nodes of the same group will preserve our rule of sign allocation that confines positive links only within, but not across, groups. 

\begin{figure}
\centering
\includegraphics[scale=0.32]{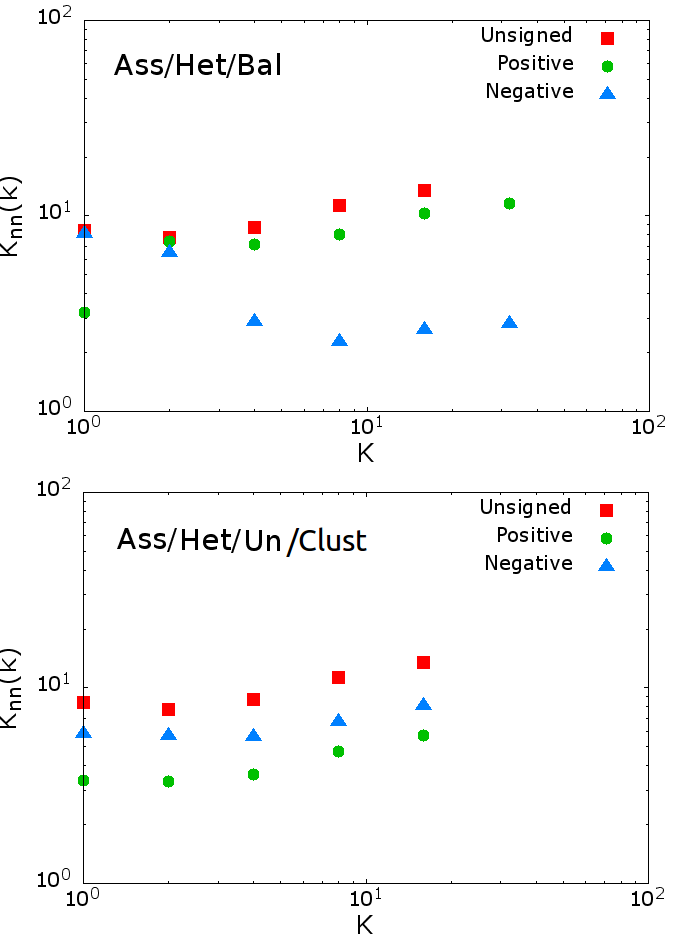}
\caption{\textbf{Positive and negative subnetworks obtained from an assortative unsigned network in the case of three groups}. The unsigned assortative network was obtained as in Fig.\ref{ass_net}. Panel (a) reports the different mixing patterns for the positive and negative subnetworks obtained under the conditions of structural balance and heterogeneous allocation of nodes into three groups. Panel (b) reports results obtained for a network with groups of unequal size and that is clusterable (Clust) but unbalanced. In all panels, data were logarithmically binned.}
\label{knn_three_groups}
\end{figure}

The trend of $K_{nn}(k)$ for the case of three groups is similar to the one obtained with two groups. Fig.\ref{knn_three_groups} reports the value of $K_{nn}(k)$ for the unsigned network and signed subnetworks under the joint conditions of assortative unsigned network, structural balance, and uneven allocation of nodes into three groups (i.e., condition 2 above). As was the case with the two groups, the negative subnetwork, unlike the positive one, is characterized by a disassortative mixing pattern. As before, these opposite trends in mixing patterns do not emerge under all the other combinations of conditions, and in particular when networks are clusterable yet unbalanced \cite{Davis1967}. Results thus suggest that clusterability is not a substitute for balance: networks that contain all-negative triangles connecting nodes from distinct groups do not display correlation patterns that differ from those obtained from unbalanced networks. 

Table \ref{new_tab} further corroborates the results from Fig.\ref{knn_three_groups}. The Table reports the values of the correlation coefficient $r$ of the degrees of connected nodes in the unsigned network and the signed subnetworks. Just as in the case of two groups, the mixing pattern of the negative subnetwork differs from the patterns of the unsigned network and positive subnetwork only when the unsigned network is assortative, the signed network is balanced, and the three groups differ in size. Indeed under these conditions, the correlation coefficient is negative for the negative subnetwork, while it remains positive for the positive one.

\begin{table}
\centering
\begin{tabular}{|l|c|c|}
\hline
\multicolumn{3}{ |c| }{\textbf{The case of three groups}} \\
\hline 
	\textbf{Conditions}&\textbf{Dis. Unsigned Network} &\textbf{Ass. Unsigned Network}\\
\hline
					&$r^u=-0.09413$                & $r^u=0.16249$     \\
\textbf{Het/Bal}    &$r^+=-0.08933$                & $r^+=0.19226$     \\
                &$r^-=-0.11434$                & $r^-=-0.2404$     \\
                \hline
                &$r^u=-0.09413$                & $r^u=0.16249$     \\
\textbf{Hom/Bal}    &$r^+=-0.09234$                & $r^+=0.15685$     \\
                &$r^-=-0.09535$                & $r^-=0.14271$     \\
                \hline
                &$r^u=-0.09413$                & $r^u=0.16249$     \\
\textbf{Het/Un}    &$r^+=-0.092480$                & $r^+=0.15243$     \\
                &$r^-=-0.095247$                & $r^-=0.07686$     \\
                \hline
                &$r^u=-0.09413$                & $r^u=0.16249$     \\
\textbf{Het/Un/Clust}    &$r^+=-0.08364$          & $r^+=0.14912$     \\
                &$r^-=-0.09524$                & $r^-=0.12176$     \\
                \hline
                &$r^u=-0.09413$                & $r^u=0.16249$     \\
\textbf{Hom/Un}    &$r^+=-0.09338$                & $r^+=0.13252$     \\
                &$r^-=-0.09395$                & $r^-=0.12230$     \\ 
                \hline
                &$r^u=-0.09413$                & $r^u=0.16249$     \\
\textbf{Hom/Un/Clust}    &$r^+=-0.093371$                & $r^+=0.18380$     \\
                &$r^-=-0.094508$                & $r^-=0.10900$     \\        						
\hline 
\end{tabular}
\caption{\textbf{Values of the correlation coefficient $r$ for the case of a network that splits into three groups.} The coefficient was calculated for each of the four combinations of structural balance (Bal) and unbalance (Un), and even (Hom) and uneven (Het) group size. Under each of the four combinations, the coefficient was calculated distinctively for each of the two cases in which the unsigned global network is assortative (and obtained through the copying model) and disassortative (and obtained with the fitness model). The variation in sign of the correlation coefficient between positive and negative subnetworks occurs only when the unsigned network is assortative, the signed network is balanced, and the three groups are of unequal size such that one dominates the other two. The coefficient was also evaluated for the cases in which the network is clusterable (Clust) but unbalanced. Results are in qualitative agreement with the values obtained when the network is unbalanced and unclusterable.}
\label{new_tab}
\end{table}

The analysis of three groups can easily be generalized to settings with any number $g$ of mutually exclusive groups. We can identify three possible cases: (i) all $g$ groups are of equal size; (ii) there is one group that dominates all other $g-1$ groups; and (iii) two or more equally sized groups dominate the remaining ones. Just as with three groups, case (iii) can be reduced to the case in which there are either two or $g$ equally sized groups, depending on the difference in size between the dominant and dominated groups. The same results obtained with three groups can be extended to this general setting: signed subnetworks display different mixing patterns only when there is one dominant group, regardless of the number of dominated groups and their comparative size.  

\section{Conclusions}
\label{last}
Our study was prompted by the empirical analysis of two signed social networks and by the observation that their mixing patterns by degree vary depending on the sign of the link. In particular, our findings indicated that negative subnetworks are characterized by disassortative patterns, in sharp contrast with their corresponding unsigned networks and the positive subnetworks. The emergence of opposite trends of mixing patterns seems to be at variance with the widely accepted belief that social networks are predominantly assortative, possibly as a result of their tendency to organize themselves into communities \cite{Newman2003}. Because both the positive and negative subnetworks have an underlying community structure, it follows that the social nature of links and the partition of nodes into communities are not, in themselves, a sufficient reason that explains why some observed social networks exhibit positive degree correlations. Our results indeed seem to suggest that the pattern of such correlations depends on the sign of the links between nodes, and thus ultimately on the type of the social relationship between individuals.  

To study the relation between sign of links and mixing patterns, we proposed a class of simple models in which nodes split into two mutually exclusive groups. We began our study with the simple case of unsigned random uncorrelated networks, and then extended the analysis by also investigating unsigned correlated networks with power-law degree distributions, and cases in which the network is organized into three or more groups. Upon attribution of signs to the links of an originally unsigned network, two distinct signed subnetworks could be extracted, each including only links with a positive or negative sign. The comparative assessment of the degree correlations in these subnetworks suggested that, when the signed network is structurally balanced and the groups differ in size, the negative subnetwork is \textit{always} characterized by a disassortative pattern, regardless of the correlation patterns displayed by the positive subnetwork and the corresponding unsigned network. In particular, under the combined conditions of structural balance and uneven group size, the correlation patterns of the two signed subnetworks differ when the unsigned network is either uncorrelated or assortative. In either case, the positive subnetwork is assortative, while the negative one is disassortative. In particular, the case of networks that split into three or more mutually exclusive groups suggested that clusterability is not a substitute for balance: when networks are clusterable but unbalanced, both signed subnetworks display the same degree correlations as the one in the corresponding unsigned network. 

By identifying the conditions under which degree correlations vary depending on the sign of the links, this study suggests that ignoring the sign would result in a loss of information on the structural properties of the network that would simply remain hidden in the unsigned network. Moreover, our findings indicate that assortativity, often regarded as a characteristic signature of most social networks, can be justified not simply by the social character of these networks, but more precisely by the positive nature of the social relationships they embody. Indeed the broad category of social networks typically subsumes a variety of relationships and interactions that are often difficult to disambiguate and may, as  result, intermingle with each other and remain confounded in one single type of connection. In such cases, detecting assortativity in a network may simply indicate either that the nature of the social relationships was ignored or that their positive components outweigh the negative ones. Conversely, disassortativity may indicate that the unsigned network is in itself disassortative or that the negative components of the links outweigh the positive ones. Finally, a lack of degree correlations may result simply from an unsigned uncorrelated network or from cases in which the positive and negative components of the relationships compensate each other out. 

The model here proposed can also account for other distinctive features of social networks. For instance, the signed subnetworks extracted from each of the two social networks here analyzed differ not only in mixing patterns, but also in clustering \cite{Szell2010a}. The positive subnetworks show a higher value of the clustering coefficient than the negative ones; moreover, in one case clustering is higher, whilst in the other is lower than would be expected by chance. Our model does indeed reproduce this property. As a result of the imposed structural balance, the negative subnetworks have a tree-like structure and are not allowed to contain closed triads, unlike the positive subnetworks in which the closure of triads does not affect balance. In this case, the difference in clustering results from our rule of sign attribution. 

Our model can be further extended to investigate degree correlations across subnetworks, namely the pairing between nodes' positive and negative degrees, and between nodes' positive (negative) degrees and their neighbors' negative (positive) degrees. More generally, our analysis can be regarded as a platform for further studies of mixing patterns in complex networks. If degree correlations vary according to the sign and nature of the connections, this study suggests that the sign of the links could, in principle, be inferred simply from the analysis of the structural properties of a network. From this perspective, our findings can help inspire the development of a quantitive measure for uncovering the hidden sign of the links from the type of mixing patterns exhibited by a network. This would prove to be useful especially in cases where the sign of links could not be assessed directly or it would be too costly to do so. For instance, gauging the collaborative or competitive properties of the relationships within and between organizations is typically constrained by a number of biases originating from the subjective, multiplex and complex nature of such relationships. These biases, however, can easily be overcome when the sign and nature of the relationships can be extracted directly from the degree correlations of the intra- and inter-organizational networks.  

\section*{Acknowledgement}

The authors acknowledge useful discussions with V. Loreto and V. D. P. Servedio.

\bibliographystyle{plain}

\end{document}